\documentclass[twocolumn,twoside,slac_two]{revtex4}
\usepackage{graphicx}
\usepackage{fancyhdr}
\newcommand{\G}{\Gamma}

\newcommand{\sT}{\sigma_{\rm T}}

\newcommand{\g}{\gamma}
\newcommand{\gp}{\gamma^{\prime}}

\newcommand{\up}{u^\prime}

\newcommand{\psim}{\lower.5ex\hbox{$\; \buildrel \propto \over\sim \;$}}
\newcommand{\lbar}{\lower.0ex\hbox{$\; \buildrel
{\lower0.0ex \hbox{-}} \over\lambda  \;$}}

\newcommand{\ph}{\mathrm{ph}}
\newcommand{\cm}{\mathrm{cm}}

\newcommand{\eV}{\mathrm{eV}}

\newcommand{\MeV}{\mathrm{MeV}}
\newcommand{\GeV}{\mathrm{GeV}}

\newcommand{\s}{\mathrm{s}}

\newcommand{\Hz}{\mathrm{Hz}}
\newcommand{\pc}{\mathrm{pc}}

\newcommand{\Gauss}{\mathrm{G}}

\begin{document}

\title{Blazars in Context in the {\em Fermi} Era}


\author{Justin D.\ Finke}
\affiliation{Space Science Division Code 7653, US Naval Research Laboratory, 4555 Overlook Ave. SW, 
Washington, DC 20375, USA}
\author{on behalf of the Fermi-LAT Collaboration}

\begin{abstract}
Blazars are the most plentiful $\g$-ray source at GeV energies, and
despite detailed study, there is much that is not known about these
sources.  In this review I explore some recent results on blazars,
including the controversy of the ``blazar sequence'', the curvature in
the LAT spectra, and the location along the jet of the $\g$-ray
emitting region.  I conclude with a discussion of alternative modeling
possibilities.
\end{abstract}

\maketitle

\thispagestyle{fancy}


\section{Introduction}

The second {\em Fermi} Large Area Telescope (LAT) catalog
\citep{nolan12_2fgl} contains 1,298 identified or associated sources,
of which 84\% are Active Galactic Nuclei (AGN) of some flavor or
another, mostly blazars.  Of the 575 unidentified sources in this
catalog, 27\% have since been associated with blazars based on
analysis of their infrared (IR) colors, as observed by the {\em Wide
Field Infrared Survey Explorer (WISE)} \citep{massaro12_IR}.  Blazars
dominate the $\gamma$-ray sky in terms of sheer number of sources.

\subsection{Basic Blazar Physics}

Blazars are thought to be powered by accretion onto supermassive
($M\sim 10^6-10^9\ M_{\odot}$) black holes at the center of what seem
to be almost entirely elliptical galaxies
\citep[e.g.,][]{bahcall97,boyce98,urry00}.  Jets are produced
perpendicular to the accretion disk probably through magnetic fields
wound up by the spin of the black hole \citep{blandford77}. The jets
are closely aligned to our line of sight, the defining property of
blazars.

The jets move at speeds close to the speed of light, $c$, with Lorentz
factors $\Gamma = (1-\beta^2)^{-1/2}\sim 10$, where the jet speed
$v=\beta c$.  These high jet speeds can be inferred from several
pieces of evidence: their extreme radio surface brightnesses that
would require extreme energy densities if produced by synchrotron from
stationary sources \citep{jones74_1,jones74_2}; the superluminal
apparent speeds ($v_{app}=\beta_{app}c$) of components seen with radio 
very long baseline interferometry \citep[VLBI; e.g.,][]{lister09}; and
the detection of rapid $\g$-ray flares implies the source must be
moving with $\G \gg 1$ to avoid $\g\g$ attenuation
\citep[e.g.,][]{dondi95}.  With a small angle, $\theta$, to the line
of sight they have Doppler factors given by $\delta =
[\G(1-\beta\cos\theta)]^{-1}$.  The observed $\nu F_\nu$ synchrotron
flux ($f_{sy}$) of a rapidly moving source compared to what its flux
would be if it were stationary ($f_{sy}^{\prime}$) is given by
$f_{sy}=\delta^4 f_{sy}^{\prime}$.  A blazar jet with fiducial values
$\G= 10$ and $\theta=1/\G$ will have $\delta = 10$ and so will be
$10^4$ times brighter than what it would be if it were stationary.
The radiation is said to be beamed in the direction of the jet's
motion, and this accounts for the extreme brightness of blazars.

The rapid variability observed in blazars at all wavelengths, from GHz
radio frequencies to TeV $\g$-rays, implies emission from a compact
region.  If a compact region of plasma (the ``blob'') is assumed to be
a sphere with radius $R^\prime$ in the frame co-moving with the blob,
then the variability timescale ($t_v$; the approximate time it takes
the flux to double) and light travel-time arguments give the
constraint
\begin{eqnarray}
\label{size}
R^\prime \le \delta c t_v/(1+z) = 
3\times10^{15}\ \delta_1 t_{v,4} (1+z)^{-1}\ \cm\ .
\end{eqnarray}
Here and everywhere, primed quantities refer to the co-moving frame.
I have used the notation that $A_x = 10^x A$ and all variables are in
Gaussian/cgs units unless otherwise stated.  The VLBI imaging of
blazar and jets often reveal individual knots
\citep[e.g.][]{piner04,lister09}, further evidence that jets consist
of discreet components.

Electrons are accelerated, probably by shocks internal to the jet, to
form power-law distributions, $N(\gp)\propto \g^{\prime -p}$.  In a
magnetic field these electrons emit synchrotron radiation, which
almost certainly is responsible for the low-energy emission in
blazars, peaking in the infrared through X-ray.  The $\g$-ray emission
from blazars is less clear but probably originates from Compton
scattering either of the synchrotron radiation (synchrotron
self-Compton or SSC) or some external radiation field (external
Compton or EC).  The external radiation field could be from a thermal
accretion disk, a broad line region (BLR), or a dust torus.  It is
also possible there could be a $\g$-ray component from emission by
protons accelerated in the jet as well.  Both leptonic and hadronic
emission models in blazars are reviewed by \citet{boett07,boett12}.

\subsection{Classification}

Blazars are sub-divided as Flat Spectrum Radio Quasars (FSRQs) and BL
Lacertae objects based on their optical spectrum, with sources with
weak or absent broad emission lines being BL Lacs, and those with
stronger broad emission lines being FSRQs.  Blazars are further
classified based on $\nu_{pk}^{sy}$, the frequency of their
synchrotron peak in a $\nu F_{\nu}$ representation.  Most recently,
they were classified as low synchrotron peaked (LSP) if $\nu_{pk}^{sy}
< 10^{14}\ \Hz$, intermediate synchrotron peaked (ISP) if $10^{14}\
\Hz<\nu_{pk}^{sy}<10^{15}\ \Hz$, or high synchrotron peaked (HSP) if
$10^{15}<\nu_{pk}^{sy}$ by \citet{abdo10_sed}.  Almost all FSRQs are
LSPs \citep{ackermann11_2lac}.  BL Lacs are generally thought to be
the aligned counterpart to FR Is, while FSRQs are generally thought to
be the aligned counterpart to FR IIs \citep[e.g.,][]{urry95}, although
some exceptions exist \citep[e.g.,][]{landt04}.

\section{Blazar Sequence}

\subsection{The Origin of the Sequence}
\label{sequence_origin}

One of the great accomplishments of twentieth century astrophysics is
the understanding of stars.  We now understand their power source, how
much radiation they produce and their spectra, how this depends on
their mass and chemical composition, and how it evolves with time.  It
is worth taking the time to think about the question: How is it that
we understand stars so well, yet we understand blazars so poorly?  Why
do we not have a good understanding of how blazars' emission and
spectra depend on fundamental parameters (black hole mass, black hole
spin, or other parameters), how they evolve with time, and so forth.

Stars are isotropic emitters, and appear mostly the same no matter
which direction one is looking at them.  For blazars, this is
obviously not the case.  Stars tend to have relatively constant emission
on human timescales, or, if they are variable, the variability is
predictable (e.g., Cepheid variables or RR Lyrae stars).  Blazars are
highly variable at all wavelengths across the electromagnetic spectrum
on time scales as short as hours or even minutes
\citep[e.g.,][]{aharonian07_2155}, and the variability is apparently
stochastic.  Globular clusters played an important role in the
understanding of stars, since one can safely assume that all of the
stars in the cluster have been created at about the same time.  There
is no such similar method for figuring out the relative ages of
blazars.  Finally, one can determine the composition, temperature, and
density of stellar photospheres from their optical spectra; as the
jets of blazars are fully-ionized, spectral lines are not expected,
and they have no similar diagnostic.

One of the most useful tools in stellar astrophysics is the
Hertzsprung-Russell diagram, which describes the luminosity to the
optical spectral type (related to temperature and color) of stars and
includes the very prominent main sequence, on which stars spend a
large fraction of their lifetimes.  This diagram has led to enormous
success in the understanding of stars, so that one is greatly tempted
to find a similar diagram for blazars.  The possible discovery of a
``blazar main sequence'' or ``blazar sequence'' was made by
\citet{fossati98}, combining three samples of blazars: a sample of
FSRQs \citep[from the 2 Jy sample of ][]{wall85}, a radio-selected
sample of BL Lacs \citep[from the 1 Jy sample of][]{kuhr81}, and an
X-ray selected sample of BL Lacs \citep[from the Einstein Slew
Survey;][]{elvis92}.  They found three parameters that appeared to be
well-correlated with the peak of the blazar synchrotron component: the
5 GHz radio luminosity, the luminosity at the peak of the synchrotron
component, and the ``$\gamma$-ray dominance'', i.e., the ratio of of
the $\g$-ray luminosity (as measured by EGRET) and the peak luminosity
of the synchrotron component.  Could one or all of these sequences
hold the same place in blazar phenomenology that the stellar main
sequence holds in stellar phenomenology?

\citet{ghisellini98} provided a physical explanation for the
correlations, or sequence, found by \citet{fossati98}.  For nonthermal
electrons accelerated as power-laws and allowed to escape a region of
size $R^\prime$ and cool through synchrotron and Compton losses, a
``cooling break'' will be found in the electron distribution at
electron Lorentz factor given by
\begin{eqnarray}
\label{gc}
\g_c^\prime = \frac{ 3 m_e c^2 }{ 4 c \sT u_{tot}^{\prime} t^\prime_{esc} }\ ,
\end{eqnarray}
where $m_e=9.1\times10^{28}$\ g is the electron mass,
$\sT=6.65\times10^{-25}\ \cm$ is the Thomson cross section,
$t^\prime_{esc}\cong R^\prime/c$ is the escape timescale, and
$u_{tot}^\prime$ is the total energy density is the frame of the
relativistic blob, given by the sum of the Poynting flux
($u_B^\prime$), synchrotron ($u_{sy}^\prime$), and external radiation
field ($u^\prime_{ext} \cong \G^2 u_{ext}$) energy densities.  Note that
all primed quantities are in the frame co-moving with the jet blob.
The cooling Lorentz factor $\gp_c$ will be associated with a peak in
the synchrotron spectrum of the source in a $\nu F_\nu$ representation
observed at frequency
\begin{eqnarray}
\label{nupk}
\nu_{pk}^{sy} = 3.7\times10^6\ \g_c^{\prime 2}\ \left(\frac{B}{\Gauss}\right)\ 
\frac{\delta}{1+z}\ \Hz
%\\ \nonumber
%= 3.6\times10^{14}\ \Hz\ B_0\ \delta_{1}\ (1+z)^{-1}
%\ R_{15.5}^{\prime -2}\ u_{tot,-1}^{\prime -2}\ .
\end{eqnarray}
\citep[e.g.,][]{tavecchio98} where $B$ is the magnetic field in the
blob.  For objects that have weak external radiation fields so that
$\up_B \gg \up_{ext}$, and neglecting $\up_{sy}$, then Equations
(\ref{gc}) and (\ref{nupk}) give
\begin{eqnarray}
\nu_{pk}^{sy} \cong 2.2\times10^{15}\ B_0^{-3}\ \delta_1\ (1+z)^{-1}\ 
\ R_{15.5}^{\prime -2}\ \Hz\ ,
\end{eqnarray}
where I have chosen fiducial values for all quantities.  These objects
will be HSPs.  Objects with a strong external radiation fields from
the broad line region (BLR) which dominate over $\up_B$ and
$\up_{sy}$, will have peak synchrotron frequencies given by
\begin{eqnarray}
\nu_{pk}^{sy} \cong 3.2\times10^{12}\ \ B_0\ \delta_1^{-3}\ (1+z)^{-1}\ 
\ R_{15.5}^{\prime -2}\ u_{ext,-2}^{-2}\ \Hz
\end{eqnarray}
where I assumed that $\delta = \Gamma$.  These objects will be LSPs.
It turns out that so far all blazars with high synchrotron peaks are
BL Lacs (without strong broad emission lines by definition), while
FSRQs with strong emission lines are almost entirely LSPs.  Note
however, that there are a significant number of BL Lacs which are
LSPs.  Objects with stronger line emission would also be expected to
have greater $\g$-ray dominances, due to scattering of the external
radiation field.  \citet{ghisellini98} thus predicted a sequence of
blazars, from low power, high peaked, low $\g$-ray dominance, lineless
objects, and as the external radiation field increases, to low peaked,
high $\g$-ray dominance objects with strong broad emission lines.
\citet{boett02_seq} suggested the ``blazar sequence'' is evolutionary,
with FSRQs being young objects, and as the circum-nuclear material
accretes, the the broad emission lines decrease, and the accretion
rate decreases, and the sources become older BL Lac objects.

However, the correlations found by \citet{fossati98} have not always
been found in subsequent studies \citep{padovani03,nieppola06}
although they have in others \citep[e.g.,][]{chen11,finke13}.
Furthermore, an alternative explanation was provided by
\citet{giommi02,giommi05,giommi12_selection}, In their scenario, the
sequence is a result of a selection effect: luminous blazars with high
synchrotron peaks will have their spectral lines totally swamped by
the nonthermal continuum, making a redshift measurement impossible.
Without a redshift, it is not possible to determine their
luminosities, and so they are not included in statistical tests
between luminosity and $\nu_{pk}^{sy}$.

What is the explanation for the blazar sequence?  Is it a physical
effect \citep{ghisellini98} or a selection effect \citep{giommi02}?

\subsection{More Recent Work}

\begin{figure}
\vspace{3.mm}
\includegraphics[width=60mm,angle=270]{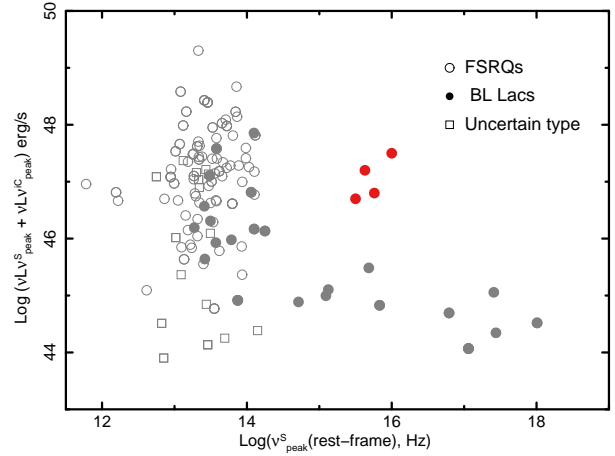}
\caption{ The sum $L_{pk}^{sy} + L_{pk}^{C}$ versus $\nu_{pk}^{sy}$
for a number of blazars.  \citet{rau12} found the redshifts for the
four luminous, high-peaked, high $z$ objects shown in red.  Figure
taken from \citet{padovani12}.  }
\label{padovani_fig}
\end{figure}
\vspace{2.2mm}

\begin{figure}
\vspace{4.mm}
\includegraphics[width=75mm]{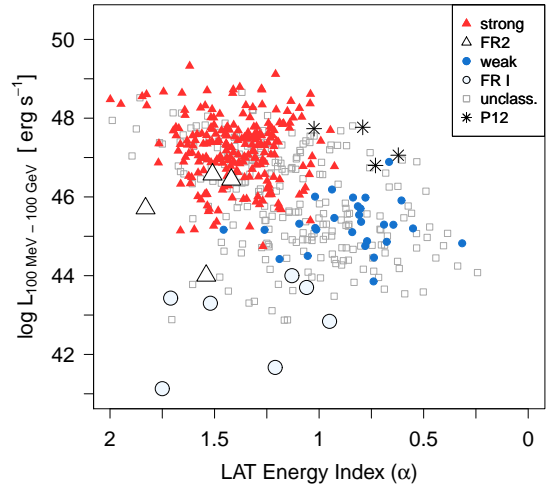}
\caption{ The LAT $\gamma$-ray luminosity versus LAT spectral energy
index ($\alpha = \Gamma_\g - 1$) from \citet{meyer12}.  FSRQ sources
are shown in red, BL Lacs are shown in blue, and the sources from
\citet{rau12} and \citet{padovani12} are shown as asterisks.  }
\label{meyer12_fig}
\end{figure}

\citet{rau12} have constrained the redshifts of a number of high $z$
BL Lacs.  Four of these do seem to have high $\nu_{pk}^{sy}$ and are
very luminous \citep[see Fig.\ \ref{padovani_fig};][]{padovani12}.
This would seem to support the argument that the blazar sequence is
the result of a selection, rather than physical, effect.  In the {\em
Fermi} era, however, it is possible to look at not just the
synchrotron component, but also the $\g$-ray component, presumably the
result of Compton scattering.  Both \citet{meyer12} and
\citet{ghisellini12} pointed out that these four sources are not out
of the ordinary on a $\g$-ray ``blazar sequence, '' where one plots
the LAT spectral index, $\G_\g$ (a proxy for the peak of the $\g$-ray
component) and the LAT $\g$-ray luminosity, and they are perfectly
consistent with other LAT $\g$-ray sources (see Fig.\
\ref{meyer12_fig}).  However, it is certainly possible that in the
future, as more redshifts are measured and constrained, sources with
high $L_\g$ and low $\G_\g$ will be found.

\vspace{2.2mm}
%\clearpage

\begin{figure}
\vspace{4.mm}
\includegraphics[width=75mm]{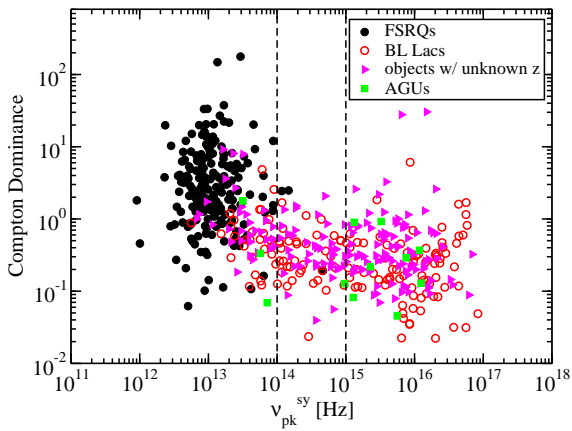}
\caption{Compton dominance (i.e., $L_{pk}^C/L_{pk}^{sy}$) versus peak
synchrotron frequency.  Filled circles represent FSRQs, empty circles
represent BL Lacs, and filled squares represent objects which do not
have an unambiguous classification.  Rightward-pointing triangles
represent BL Lacs with unknown redshifts, for which $\nu_{pk}^{sy}$ is
a lower limit.  Figure taken from \citet{finke13}.}
\label{CD}
\end{figure}
\vspace{1.2mm}
%\clearpage

An often overlooked part of the blazar sequence as found by
\citet{fossati98} is the $\g$-ray dominance, i.e., the ratio of the
$L_\g$ to the peak synchrotron luminosity ($L_{pk}^{sy}$).  This and a
similar quantity, the Compton dominance, $A_C \equiv L_{pk}^C /
L_{pk}^{sy}$ (where $L_{pk}^C$ is the luminosity at the Compton peak)
are redshift-independent.  Also, $\nu_{pk}^{sy}$ is only weakly
dependent on redshift, by a factor $(1+z)$, i.e., a factor of a few.
A plot of $A_C$ versus $\nu_{pk}^{sy}$ is shown in Fig.\ \ref{CD},
from a subset of sources in the second LAT AGN catalog
\citep{ackermann11_2lac}, including sources which do not have known
redshifts.  It is clear a correlation exists, and this is confirmed
with the Spearman and Kendall tests \citep{finke13}.  It seems that
this aspect of the blazar sequence has a physical origin, and is not
the result of a selection effect.  In the future, the luminosity-peak
frequency relations could be improved with new redshift measurements
and constraints \citep[e.g.,][]{shaw13}.  Then it should be possible
to determine if these aspects of the sequence are physical as well.

As an alternative to the physical scenario described by
\citet{ghisellini98} and in \S\ \ref{sequence_origin}, \citet{meyer11}
proposed another physical scenario, based on updated data from a
number of sources.  In their scenario, the difference between BL Lacs
and FSRQs is the former have jet structure with velocity (or Lorentz
factor) gradients, either perpendicular or parallel to the direction
of motion.  FSRQs, according to \citet{meyer11}, do not have these
gradients; they have a single Lorentz factor for the entire jet, or at
least the radiatively important parts.  There is indeed ample evidence
for different Lorentz factors in BL Lacs and FRIs
\citep[e.g.,][]{chiaberge00,chiaberge01,abdo10_cena}.  The lack of
$\g$-ray detected FRIIs hints that FRIIs/FSRQs do not share this jet
structure \citep{grandi12}, however, see \citet{boett09_decel} for
evidence of jet deceleration in an FSRQ.

\section{Curvature in LAT Spectra}
\label{curvature}

After the launch of {\em Fermi}, while the spacecraft was still in its
post-launch commissioning and checkout phase, the FSRQ 3C~454.3 was
detected by the LAT in an extreme bright state \citep{tosti08_atel}.
The source reached a flux of $F(>100\ \MeV) > 10^{-5}\ \ph\ \cm^{-2}\
\s^{-1}$ and its spectrum showed an obvious curvature (i.e., a
deviation from a single power-law), which was best-fit by a broken
power-law \citep{abdo09_3c454.3} with break energy $\sim 2$\ GeV.
This source flared on several more occasions
\citep{ackermann10_3c454.3,abdo11_3c454.3}, always exhibiting a
spectral break during bright states.  The energy of the break varied
by no more than a factor of $\sim 3$, while the flux varied by as much
as a factor of 10 \citep{abdo11_3c454.3}.  This spectral curvature has
been found in other blazars as well, although a broken power-law is
not always preferred over a log-parabola fit, which has one less free
parameter \citep{abdo10_latsed}.  The cause of the break is not clear
but there are several possible explanations.

{\em A combination of several scattering components.}
\citet{finke10_3c454} noted that, based on the shape of the optical
and $\g$-ray spectra, the Compton scattering of more than one seed
photon source was needed to explain the overall spectral energy 
distribution (SED) of 3C~454.3.  The
particularly soft spectra above the break requires that this
scattering be done in the Klein-Nishina (KN) regime.  This model
requires that scattering occur within the BLR, and a wind model for
the BLR in order to explain the relative stability of the break
energy.

{\em Photoabsorption of $\g$-rays with BLR photons}.  \citet{pout10}
\citep[see also][]{stern11} pointed out that He II Ly$\alpha$ and
recombination photons are at the right energy (54.4 eV and 40.8 eV,
respectively) to absorb $\g$-ray photons at $\sim 5$\ GeV, about the
same energy as the spectral breaks observed.  This model would also
require the $\g$-ray emitting region to be within the BLR.

{\em Compton scattering of BLR Ly$\alpha$ photons}.  For the
scattering of Ly$\alpha$ photons ($E_*=10.2$\ eV), the KN regime will
emerge at energies above
\begin{eqnarray}
E_{KN} \approx 1.2\ (E_*/10.2\ \eV)^{-1}\ \GeV\ ,
\end{eqnarray}
approximately in agreement with the observed break energy
\citep{ackermann10_3c454.3}.  Fits with this model using power-law
electron distributions failed to reproduce the observed LAT spectra
\citep{ackermann10_3c454.3}; however, fits using a log-parabola
electron distribution were able to reproduce the $\g$-rays
\citep{cerruti12}.  Naturally, this model would also require the
$\g$-ray emitting region to be within the BLR.

\begin{figure}
\vspace{3.mm} 
\includegraphics[width=80mm]{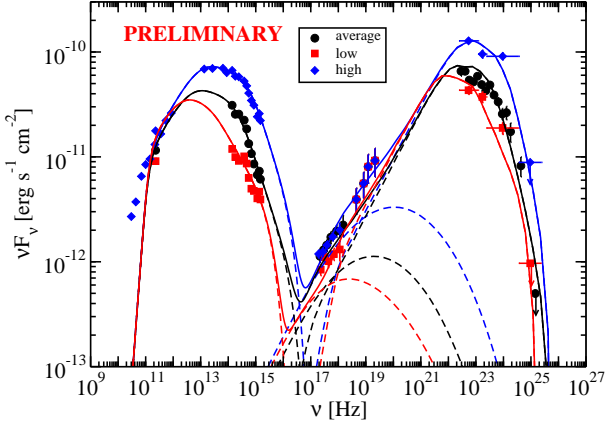}
\caption{SED of the FSRQ PKS~0537$-$441 from \citet{dammando13}.  The
spectral curvature in the IR/optical in the high state could indicate
the cause of the $\g$-ray curvature is the result of curvature in the
electron distribution.  }
\label{0537sed}
\end{figure}
\vspace{2.2mm}

{\em Curvature in the electron distribution}.  This is the explanation
originally favored by \citet{abdo09_3c454.3}.  If there is curvature
in the electron distribution that produces the $\g$-rays, presumably
from Compton scattering, this would naturally be reflected in the LAT
spectrum as well.  In this scenario, one would expect the curvature in
the electron distribution to cause a curvature in the synchrotron
emission from the same electrons, which would appear in the
IR/optical.  Indeed, observations of PKS 0537$-$441 do show this
curvature \citep[][see Fig.\ \ref{0537sed}]{dammando13}.  This
explanation would not require scattering to take place in the BLR, as
dust torus photons could be the seed photon source for scattering.
This scenario begs the question: what is the cause of the break in the
electron distribution?

\section{Location of the $\gamma$-ray Emitting Region}
\label{region}

Many of the scenarios described in \S\ \ref{curvature} require the
$\g$-rays to be produced within the BLR.  However, it is not clear
that this is the case.  Optical and $\g$-ray flares are often
associated with the rotation of polarization angles, the slow increase
in radio flux, and the ejection of superluminal components from the
core as seen at 43 GHz \citep[e.g.,][]{marscher08,marscher10}.
According to \citet{marscher12}, 2/3 of $\g$-ray flares are associated
with the ejection of a superluminal component, indicating the $\g$-ray
flares are coincident with the 43 GHz core.  There are two arguments
that the 43 GHz core is located at, and the $\g$-ray flares originate
from, $>$ a few pc, outside the BLR. (1) Using the observed radius of
the 43 GHz core ($R_{core}$), and assuming a conical jet with a half
opening angle $\alpha$ \citep[measured, e.g., by ][]{jorstad05}, one
can determine the distance of the core from the base of the jet,
$r=R_{core}/\alpha$ \citep{agudo11}.  (2) The $\g$-ray flares occur in
the same region as the much slower radio outbursts and/or polarization
angle swings lasting 10s of days \citep[$\Delta
t$;][]{marscher10,orienti13}.  The distance associated with the light
travel time of these radio outbursts or polarization swings assuming
$\theta\ll 1$ is
\begin{eqnarray}
r \ge 1.0\ \Delta t_{6}\ \delta_1\Gamma_1\ (1+z)^{-1}\ \pc\ .
\end{eqnarray}

On the other hand, the rapid $\g$-ray variability observed in blazars
such as 3C~454.3 \citep[$\sim 3$\ hours;][]{tavecchio10},
PKS~1510$-$089 \citep[$\sim 1$\ hour;][]{brown13,saito13}, 4C~21.35
\citep[$\sim 10$\ minutes; source also known as PKS
1222+21;][]{aleksic11}, and PKS~2155$-$304 \citep[$\sim 5$\
  minutes;][]{aharonian07_2155} limits the size of the emitting region
by Equation (\ref{size}).  If the emitting region takes up the entire
cross section of a conical jet, then it should be at a distance
\begin{eqnarray}
r \le 0.1\ \delta_1\ t_{v,4}\ \alpha_{-2}^{-1}\ (1+z)^{-1}\ \pc\ 
\end{eqnarray}
from the base of the jet.  Based on scaling relations found from 
reverberation mapping, the typical BLR region for FSRQs 
is $r_{BLR}\sim 0.1\ \pc$ \citep[e.g.,][]{bentz06}.

4C~21.35 was detected to have flux-doubling timescales of $\sim 10$\
minutes, as measured by MAGIC, out to 400 GeV.  The $\g\g$ optical 
depth is
\begin{eqnarray}
\tau_{\g\g} = \int^\infty_{\max[r,r_{BLR}]} d\ell\ \sigma_{\g\g}\ u_{BLR}/E_*
\end{eqnarray}
We can estimate the $\g\g$ cross section $\sigma_{\g\g}\cong \sT/3$,
and $u_{BLR}\cong$\ constant for $r<r_{BLR}$ .  I will use $E_* =
10.2$\ eV, i.e., for Ly$\alpha$.  If $r<r_{BLR}$ then
\begin{eqnarray}
\tau_{\g\g} \cong 40\ u_{BLR,-2}\ r_{BLR,17.5} (E_{*}/10.2\ \eV)^{-1}\ ,
\end{eqnarray}
so $\g$-rays with energies above the threshold energy, about $50\
\GeV\ (E_*/10.2\ \eV)^{-1}$ will clearly not be able to escape the
BLR.

Several ways to avoid $\g\g$ attenuation have been suggested, such as
energy transport through neutron beams \citep{dermer12_21.35}, or
$\g$-ray conversion to axions \citep{tavecchio12_axion}.  Otherwise,
the $\g$-rays from this source must be produced by a small fraction of
the jet cross section at $\ge 4$\ pc from the black hole, outside the
BLR.

If the $\g$-ray emitting region is within the BLR, the seed photons
for Compton scattering are likely to be at higher energies than they
would be if they emitting region was outside the BLR, where
lower-energy dust torus photons would serve as the seed photon source.
Due to KN effects, the Compton cooling will be different in these
different cases.  \citet{dotson12} suggest that because of this
effect, detailed study of the $\g$-ray light curves could distinguish
the seed photon source energy, and hence, the location of the emitting
region.

\section{The End of the One-Zone Leptonic Model?}

One-zone leptonic models (1ZLMs), where the lower energy emission is
produced by synchrotron radiation, and the higher energy emission is
produced by Compton scattering with the same electron population (SSC
or EC) has been the standard for fitting multi-wavelength blazar SEDs.
However, lately the multi-wavelength coverage has become complete
enough that in many cases these models do not provide sufficient fits
to blazar SEDs.  These include 3C~279 \citep[LSP
FSRQ;][]{boett09_3c279} , PKS~2005$-$489 \citep[HSP BL
Lac;][]{abram11_2005}, AO~0235+164 \citep[LSP BL
Lac;][]{ackermann12_0235}, 1ES 0414+009 \citep[HSP BL Lac;][]{aliu12},
PKS~1510$-$089 \citep[LSP FSRQ;][]{nalew12_1510}, and AP Lib [LSP BL
Lac; Abramowski et al.\ 2013, in preparation].  What is the next step
in blazar modeling?  I list three broad categories of models:

{\em Multi-zone models.}  It has been known for some time that the
flat radio spectra (index $\alpha_r\approx0$) of blazars are almost
certainly explained by the superposition of several self-absorbed
components \citep{konigl81}, so these models are perhaps the most
obvious.  The hard TeV spectra seen in some blazars, such as
1ES~1101$-$232, led \citet{boett08} to suggest the TeV emission was
produced by Compton-scattering cosmic microwave background (CMB)
photons in the kpc-scale jet.  Several models have been motivated by
the contradictory clues for the location of the $\g$-ray emitting
region, as described in \S\ \ref{region}.  These typically include a
smaller region at a large distance from the black hole, with one or
more other regions accounting for the slower radio emission
\citep[e.g.,][]{marscher10_multizone,tavecchio11_21.35,nalew12_1510}.
Inhomogeneous jets have also been explored by \citet{graff08} and
\citet{joshi11}.

{\em Hadronic models.}  Blazars have long been a candidate for the
production of ultra-high energy cosmic rays (UHECRs), a hypothesis
that was recently strengthened by the correlation of UHECRs observed
by the Auger observatory with local AGN \citep[e.g.,][]{abraham07}.
This has motivated blazar emission models where the $\g$-rays come
from processes originating from protons and cosmic rays accelerated in
the jet \citep[e.g.,][]{muecke03}.  Variability in hadronic models is
difficult to model, although progress has been made recently by
\citet{boett12}.  A neutral beam model was recently proposed by
\citet{dermer12_21.35} to explain the rapidly varying very-high energy
(VHE) emission from 4C~21.35 (see \S\ \ref{region}).

{\em Intergalactic cascade models.}  If blazars are sources of
UHECRs, the particles that escape the jet could interact with the
extragalactic background light (EBL) from stars, dust, and the CMB, to
produce cascade VHE $\g$-rays \citep{essey10_1,essey10_2}.  In this
case the VHE emission would not be variable, and would be expected to
be disconnected from the rest of the SED.  This is a simple prediction
that could be used to test this hypothesis.  VHE $\g$-rays could also
be produced from VHE $\g$-rays which interact with the EBL to produce
$e^+/e^-$ pairs.  These pairs could then in turn Compton-scatter CMB
photons, producing $\g$-rays in the LAT bandpass.  This creates
another component that needs to be taken into account in spectral
modeling of blazars \citep{davezac07,tavecchio11_igmfmodel}.

The problem with alternatives to the 1ZLM is that the addition of free
parameters means that no matter what model is used, one will almost
certainly be able to adjust the parameters to fit the data.  Both
theoretical and observational advances are needed to advance our
understanding of these sources.

\begin{acknowledgments}

I would like to thank C.\ Dermer, S.\ Razzaque, and B.\ Giebels for
discussions on topics presented in this proceeding.  I would also like
to thank the conference organizers for the opportunity to speak at the
{\em Fourth Fermi Symposium}, and for organizing an interesting and
enjoyable conference.

The {\em Fermi} LAT Collaboration acknowledges support from a number
of agencies and institutes for both development and the operation of
the LAT as well as scientific data analysis. These include NASA and
DOE in the United States, CEA/Irfu and IN2P3/CNRS in France, ASI and
INFN in Italy, MEXT, KEK, and JAXA in Japan, and the K.~A.~Wallenberg
Foundation, the Swedish Research Council and the National Space Board
in Sweden. Additional support from INAF in Italy and CNES in France
for science analysis during the operations phase is also gratefully
acknowledged.

\end{acknowledgments}

\bigskip % extra skip inserted
%\begin{thebibliography}{9}   % Use for  1-9  references
%\begin{thebibliography}{99} % Use for 10-99 references

%\bibliographystyle{apj}
\bibliography{3c454.3_ref,EBL_ref,references,mypapers_ref,blazar_ref,sequence_ref,SSC_ref,LAT_ref}

%\bibitem{accelconf-ref}
%http://www.cern.ch/accelconf

%\bibitem[Other et al.(1996)]{exampl-ref}
%A.N. Other, ``A Very Interesting Paper'', EPAC'96, Sitges, June
%1996.

%\bibitem{templates-ref}
%http://www.cern.ch/accelconf/templates.html

%\end{thebibliography}

\end{document}